\documentclass[aps, prl, reprint, superscriptaddress, showpacs]{revtex4-1}
\usepackage{amsmath}
\usepackage{graphicx}
\usepackage{bm}
\usepackage{amssymb}
\usepackage[colorlinks, linkcolor=blue, citecolor=blue]{hyperref}

\begin{document}

  \title{Cross derivative: a universal and efficient method for phase transitions in classical spin models}
%  \author{Yong Chen$^1$, Kai Ji$^{2,\dagger}$, Z. Y. Xie$^{3,\ddagger}$, and J. F. Yu$^1$}
%  \email{yujifeng@hnu.edu.cn}\\
%  \affiliation{$^1$Department of Applied Physics, School of Physics and Electronics, Hunan University, Changsha 410082, China}
%  \email{qingtaoxie@ruc.edu.cn}\\
%  \affiliation{$^2$Department of Physics, Shanghai Normal University, Shanghai}
%  \email{kji@shnu.edu.cn}
%  \affiliation{$^3$Department of Physics, Renmin University of China, Beijing 100872, China}

%  \email[$* $]{ yujif121779@gmail.com}
%  \email[$\ddagger $]{ qingtaoxie@ruc.edu.cn}
%  \email[$\dagger $]{ kji@shnu.edu.cn}
  %\affiliation{$^3$Institute of Physics, Chinese Academy of Sciences, P.O. Box 603, Beijing 100190}
  %\affiliation{$^4$University of Chinese Academy of Sciences, Beijing 100049}
  %\affiliation{$^5$Collaborative Innovation Center of Quantum Matter, Beijing 100190}

    \author{Y. Chen}
    \affiliation{Department of Applied Physics, School of Physics and Electronics, Hunan University, Changsha 410082, China}

    \author{K. Ji}
      %\email{kji@shnu.edu.cn}
     \affiliation{Department of Physics, Shanghai Normal University, Shanghai 200234, China}

    \author{Z. Y. Xie}
      \email{qingtaoxie@ruc.edu.cn}
    \affiliation{Department of Physics, Renmin University of China, Beijing 100872, China}

    \author{J. F. Yu}
      \email{yujifeng@hnu.edu.cn}
      \affiliation{Department of Applied Physics, School of Physics and Electronics, Hunan University, Changsha 410082, China}

  %\date{\today}

  \begin{abstract}

    %%%By applying an external weak magnetic field, we investigate the system response, and demonstrate clearly the topological Kosterlitz-Thouless (KT) transition in the continuous $XY$ model and the $5$-state clock model on a square lattice.

    %%%The transition temperatures are calculated, and agreed well with others' predictions. Besides, the phase transition between the ordered and the quasi-ordered phases in the $5$-state clock model, exhibits quite different feature from KT scenario.

    %%%By the tensor renormalization group method based on the higher-order singular value decomposition, we have investigated the thermodynamic properties of the $5$-state clock model on the square lattice, by applying a weak external magnetic field. The system exhibits two phase transitions clearly, and the transition temperatures are determined from the peaks of the magnetization derivative of temperature, at $T_{c1}=0.9065(4)$ and $T_{c2}=0.9525(8)$ respectively, consistent with the results of Monte Carlo and other methods.

    %More importantly, these two peaks vary with the applied magnetic field in distinctly different trends, and our simulations indicate their different transition types: the upper one is a Kosterlitz-Thouless transition, but the lower one is not.

    %%%Most importantly, our calculations provide clear evidence that, the higher temperature transition is a Kosterlitz-Thouless transition; while the lower one is not, for it is dominant by the mixed topological excitations of vortices and domain walls.

    With an auxiliary weak external magnetic field, we reexamine the fundamental thermodynamic function, Gibbs free energy $F(T, h)$, to study the phase transitions in the classical spin lattice models.     A cross derivative, i.e. the second-order partial derivative of $F(T, h)$ with respect to both temperature and field, is calculated to precisely locate the critical temperature,   which also reveals the nature of a transition.
    The strategy is efficient and universal, as exemplified by the 5-state clock model, 2-dimensional (2D) and 3D Ising models, and the $XY$ model, no matter a transition is trivial or exotic with complex excitations.
More importantly, other conjugate pairs could also be integrated into a similar cross derivative if necessary, which would greatly enrich our vision and means to investigate phase transitions both theoretically and experimentally.

    %Investigating the phase transitions of the 5-state clock model on the square lattice, we rekindle the fundamental thermodynamic function, Helmholtz free energy, with an auxiliary weak magnetic field. The transition temperatures are precisely located, by calculating the second-order derivative of free energy with respect to both the magnetic field and the temperature, which also indicates the nature of the phase transitions. The method is convenient and universal, as well illustrated for the classic Ising and the $XY$ models, no matter a transition is trivial, or exotic with complex excitation(s).

    %%%For 2-dimensional (2D) classical spin systems, we investigate the free energy and its derivatives, with respect to an auxiliary weak magnetic field $h$ and the temperature $T$,  and the phase transitions can be easily detected and depicted, no matter it is of KT type or not.

  \end{abstract}

%  \pacs{05.70.Fh, 05.10.Cc, 75.10.Hk}

  \maketitle

%\section{Introduction}\label{introduction}
Phases of matter and phase transitions have always been the hot topics in the statistical and the condensed matter physics. For decades, Landau's symmetry-breaking theory was believed fully qualified to identify and describe different phases and phase transitions in-between.
As a seminal illustration of the spontaneous symmetry breaking, the classical Ising model ($Z_2$ symmetry) on the square lattice, undergoes a typical order-disorder phase transition. Its opposite extreme, the continuous $XY$ model ($U(1)$ symmetry), involves the exotic topological vortices excitation and a phase transition without symmetry broken, i.e. the so-called Kosterlitz-Thouless (KT)\cite{KT1,KT2} transition beyond Landau's theory.
Both types of transitions can be easily probed by the magnetic susceptibility, which reflects the system's response to an external magnetic field and behaves distinctively across the critical point.

One natural question is how the universality class evolves with the symmetry of the models, which arouses intensive interest in the intermediate $q$-state clock model with a finite $q$. As well known, when $q$ is no bigger than 4, it has one unique second-order phase transition; otherwise, there are two separate transitions sandwiching a critical KT phase with quasi-long-range order. So far, major debates focus on $q$ near 5, about the nature and the precise locations of the transitions.

Usually, Monte Carlo (MC), one of the principal methods for many-body problems, calculates the helicity modulus\cite{Lapilli, Baek1, Baek3, Kumano, Okabe} to characterize the KT transition. As in the continuous $XY$ model on the square lattice, it jumps abruptly from finite to zero at the critical point\cite{Minnhagen}.
For the $5$-state clock model, at the upper transition point, it behaves similarly to the $XY$ case. However, as to the lower one, inconsistent conclusions about the transition type were claimed by different groups with MC studies\cite{Kumano, Baek3}.
By proposing an extended universality class theory with MC simulations, Lapilli et al.\cite{Lapilli} even declared both transitions are not KT-type when $q\le6$, which is supported by Ref.~[\onlinecite{Hwang}] from a Fish zero analysis for $q=6$ case.

\begin{figure}[htbp]
    \begin{center}
      \includegraphics[width=0.48\textwidth,clip,angle=0]{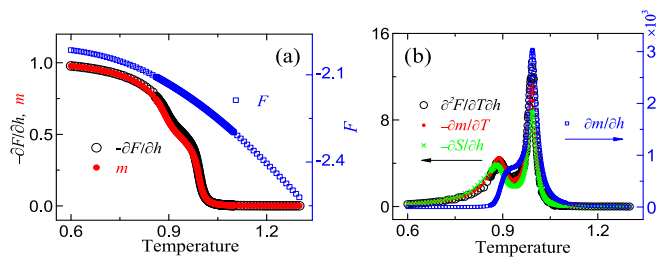}
           % Here is how to import EPS art [width=20mm,height=10mm][width=9cm,trim=0.5cm 21cm 0 2cm]
      \caption{\label{fe}(Color online) (a) Gibbs free energy $F$ (blue blank square) of 5-state clock model with a magnetic field $h=4.0 \times 10^{-5}$; comparison of $-\partial F/\partial h$ (black empty circle) and $\boldsymbol{m}$ (red filled circle); (b) magnetic susceptibility $\partial \boldsymbol{m} / \partial h$ (blue blank square), cross derivative ${\partial}^2 F / \partial T \partial h$ (black blank circle), $-\partial \boldsymbol{m} / \partial T$ (red filled circle), and $-\partial{S} / \partial h$ (green cross).}
    \end{center}
  \end{figure}

 Another powerful method, the renormalization group (RG) predicted two KT transitions early\cite{Kadanoff}, and a recent density matrix renormalization group (DMRG) study\cite{Christophe} favored this assertion by calculating the helicity modulus with relatively small system sizes.
The tensor network states, generalized from DMRG to higher dimensional strongly correlated systems, have developed rapidly and been widely used to investigate both the classical and the quantum systems. Among those, the tensor renormalizaton group method based on the higher-order singular value decomposition (abbreviated as HOTRG)\cite{Xie2}, has been successfully applied to study the 3D Ising model\cite{Xie2}, the Potts model\cite{MPQin, WShun}, and the continuous $XY$ model\cite{JFYu}. Actually, it has been also utilized to study the 5-state clock model, where the magnetic susceptibility properly describes the upper phase transition, but does not work well for the lower one\cite{Chenyong}, as also presented in Fig. \ref{fe}(b) below. Therefore, a gauge invariant factor from the fixed point tensor of the RG flow, proposed in Ref.~[\onlinecite{Gu}], was adopted to measure the degeneracy of each phase, which also precisely estimates the critical points of the 6-state clock model\cite{Jing}.

Nevertheless, some important information is still missing, e.g., the reason why the magnetic susceptibility loses its efficacy for the lower transition in this model, and the nature of the transitions in particular,
%while performs perfectly for the upper one.
%and gave some results about the phase transition\cite{Chenyong}. Initially, the magnetic susceptibility was to characterize the transition, for which did well in describing the KT transition of the continuous $XY$ model\cite{JFYu}. Indeed, it captures the higher temperature $KT$ transition fully, but poorly identifies the lower temperature one\cite{Borisenko2011,Chenyong},
%and we had to turn to another quantity, i.e. a gauge invariant factor $X$ proposed in Ref.~[\onlinecite{Wen}] to measure the degeneracy of a phase. This quantity distinguished different phases clearly, as  also found in $6$-state clock model \cite{Jing}, and the critical points can be located properly. But still, we have missed the information about the lower temperature transition nature, and especially why the magnetic susceptibility loses its effectiveness wherein, but well-behaved for the higher temperature one.
although many researches claimed both are KT transition.
A duality analysis by the conformal field theory (CFT) deemed that two transitions are KT-type, but they still have some differences\cite{Elitzur, Matsuo}. Recently, a universal entropy predicted by CFT on a Klein bottle\cite{TuPRL, TuPRB} has different values at these two critical points\cite{TuPrivate}, which is believed valuable to distinguish different CFTs.

Here, we intend to detect and clarify the nature/mechanism of the classical phase transitions, within a unified frame, by reexamining the fundamental thermodynamic function, Gibbs free energy $F$, which is intrinsically a signpost of the universal entropy increase of a spontaneous change\cite{Atkins}, and contains information about the phase transitions.
However, the free energy and its temperature derivatives, i.e. the internal energy and the specific heat, are analytically continuous without any singularity in the KT transition. 
So, besides the temperature, we propose an auxiliary parameter, a weak external magnetic field, which interacts with spin degree of freedom and competes with thermal excitations, thus provides us a convenient tool to investigate the dynamical behavior of the system.
By detailed analyses on the cross derivative of $F(T, h)$ with respect to both temperature and field, we can easily identify and precisely locate the transition points. Moreover, since the free energy is fundamental, this idea is readily applied to any classical spin system, like Ising or $XY$ model with trivial or exotic transitions. In other words, it is universal.

%Here, we propose a simple while convenient way to solve this dilemma, by just rekindling the familiar thermodynamic property, the Gibbs energy, with inclusion of a weak external magnetic field. By detailed analyses on the interplay between these two tunable factors, one can easily identify and locate the transition temperatures. What's more, this method is readily applicable for other classical systems, trivial or exotic as also illustrated below, like the Ising model or $XY$ model on a square lattice.

%\section{Model and Method}\label{ModelMethod}

First, we demonstrate this idea explicitly by the ferromagnetic $5$-state clock model with an in plane magnetic field, whose Hamiltonian is written as
%we write the Hamiltonian of the ferromagnetic $q$-state clock model with an in plane magnetic field as
  \begin{equation}
    H = - J \sum_{\left< ij \right>}\cos(\theta_{i}-\theta_{j})-h\sum_{i}\cos\theta_{i},
  \end{equation}
where $\left< ij \right>$ means summing over all nearest neighbors. $\theta_i$ is the spin angle on lattice site $i$, selected from $\theta=2\pi k/q$, $k = 0, 1, 2, \ldots, q-1$. $J$ is the nearest coupling. $h$ is the applied field in unit of $J/\mu$, and $\mu$ is the magnetic moment of each spin. Both $J$ and $\mu$ are set as 1 for convenience.

%According to the tensor network states method, the partition function of a classical statistical system with local interactions can be expressed as tensorial product of a tensor network, each residing on one lattice site. Including an external magnetic field, each local tensor is composed by a singular value decomposition of the Boltzmann factor as illustrated in Ref.~[\onlinecite{Chenyong}] in detail. 
%Based on the foregoing description, HOTRG supplies us a convenient way to compute the system partition function, or equivalently the free energy. 
%Its coarse-graining process is demonstrated in Fig. 1 of Ref.~[\onlinecite{Xie2}]. As a real-space renormalization, each step contracts the system by half. Alternating the process horizontally and vertically, one can efficiently deal with the thermodynamic limit by dozens of iterations only, overcoming the finite size problem encountered by other numerical methods.

Here, we employ HOTRG method to compute the desired physical quantities. Details about the algorithms are in Refs.~[\onlinecite{Xie2, JFYu, Chenyong}].
Its accuracy, like other RG algorithms, is subject to the number of states kept during the RG process, labelled by the bond dimension $D$. Initially, it equals $q$, then expands exponentially along the RG process. Therefore, a truncation is necessary to ensure further steps sustainable.
%In principle, bigger $D$, higher the accuracy, though not always monotonously related.

%\section{Results and Discussions}\label{ResultsDiscussions}
The free energy $F(T, h_1)$ with field $h_1$ is presented in Fig. \ref{fe}(a), wherein $-\partial F/ \partial h$ and the magnetization $\boldsymbol{m}$ are also shown.
For comparison, the quantity $-\partial F/ \partial h$ is computed directly from
%  \begin{equation}
%    -\partial F/ \partial h = -[F(h_2)-F(h_1)]/(h_2-h_1),
%  \end{equation}
$-[F(h_2)-F(h_1)]/(h_2-h_1)$
by using two close field strengths, and $\boldsymbol{m}$ is calculated by the impurity tensor algorithm\cite{Xie2, JFYu, Chenyong}.
%Because of its discrete symmetry, the normal magnetization fluctuates due to a random symmetry breaking at low temperature, and a reorganized version is usually adopted, $m=\sqrt{\langle\cos\theta\rangle^2+\langle\sin\theta\rangle^2 }$, as in Refs.[\onlinecite{Borisen1, Lapilli, Chenyong}].
One can see they agree well with each other as should do. For this model, as discussed in Ref.~[\onlinecite{Chenyong}], the magnetic susceptibility can clearly identify the upper phase transition, but not be so convenient for the lower one. As shown in Fig. \ref{fe}(b) by blue blank squares, an exponential divergence clearly labels a phase transition near $T=1.0$. Meanwhile, a broad shoulder-shape structure emerges below, indicating something happens, but not as evident as the upper one.

Instead, the cross derivative of the free energy with respect to both temperature and field, i.e. $\partial^2{F}/{\partial{T}\partial{h}}$, is able to characterize both transitions simultaneously. Clearly, as shown in Fig. \ref{fe}(b) by black blank circles, two separate sharp peaks show up.
In particular, the upper one is coincident perfectly with the susceptibility curve, for both the position and the shape, although it decays exponentially from a much smaller peak other than divergence as in the magnetic susceptibility.
The lower one, small but still obvious, locates near $T=0.90$. Here, a relatively small bond dimension $D=40$ is used just for illustration.

As verified in Fig. \ref{fe}(a), $-\partial{F}/{\partial{h}}$ is just $\boldsymbol{m}$, then the cross derivative equals the temperature derivative of the magnetization $-\partial \boldsymbol{m} / \partial T$. As both shown in Fig. \ref{fe}(b), they match up well with each other. Similarly, one can choose function $-\partial S / \partial h$, as also presented in Fig. \ref{fe}(b), because $-\partial{F}/{\partial{T}}$ is just the thermodynamic entropy $S$.
Additionally, the Maxwell relation\cite{Reichl} $\partial{S}/\partial{h}=\partial{\boldsymbol{m}}/\partial{T}$ is numerically verified by computing $S$ directly from the difference between Gibbs free energy and the internal energy, because both terms essentially spring from the cross derivative.
For numerical simplicity and convenience, we adopt the notation $-\partial \boldsymbol{m} / \partial T$ hereafter, while keeping in mind its physical origin.

Some may question the validity or the physical meaning of this cross derivative. One can imagine slicing the 3D curved surface $F(T, h)$ along $h$-axis, then performing the derivative $\partial{F}/{\partial{T}}$ for each $h$ slice, and observing its evolution along $h$-axis; or equivalently slicing $F(T, h)$ along $T$-axis and obtaining $\partial{F}/{\partial{h}}$, then investigating its evolution along $T$-axis. Thus, each captures the effects of both temperature and field, and the system dynamics can be easily deduced. This scheme may be elaborated by a formula
  \begin{equation}
    \left( \frac{\partial}{\partial T}+\frac{\partial}{\partial h}\right)^2 F = \nabla ^2 F + 2\frac{\partial^2 F}{\partial T\partial h} ,\label{formu2}
  \end{equation}
where the left part in parentheses is a linear combination of two derivative operators in the two-dimensional orthogonal space expanded by temperature and field, and the Laplacian stands for the second-order derivatives with respect to each individual parameter, i.e. the specific heat and the magnetic susceptibility respectively, while neither is adequate to characterize the system dynamics comprehensively.

  \begin{figure}[htbp]
    \begin{center}
      \includegraphics[width=0.48\textwidth,clip,angle=0]{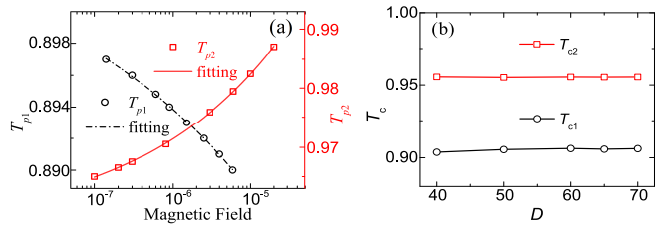}
           % Here is how to import EPS art [width=20mm,height=10mm][width=9cm,trim=0.5cm 21cm 0 2cm]
      \caption{\label{q5_Tc}(Color online) (a) Illustration of the peak positions of $-\partial \boldsymbol{m} / \partial T$ versus the magnetic fields for 5-state clock model with $D=40$, along with a power law fitting to extrapolate the transition temperature as $T_{c1}=0.9038$ and $T_{c2}=0.9557$ respectively; (b) The transition temperature versus the tensorial bond dimension $D$ to obtain the converged $T_c$ as 0.9063 and 0.9557, respectively.}
    \end{center}
  \end{figure}

%Physically, this quantity springs from the second-order partial derivative of the free energy with respect to both the temperature and the magnetic field, i.e. $\partial^2{F}/{\partial{T}\partial{h}}$.
%As also shown in Fig. \ref{fe}(b) as black blank circle, it matches up well with $-\partial \boldsymbol{m} / \partial T$ except for tiny numerical errors, for $\boldsymbol{m}$ just equals $-\partial{F}/{\partial{h}}$, and the second-order partial derivative above can properly reflect the interactive effects of the field and the temperature.
%For numerical simplicity and convenience, we then talk about $-\partial \boldsymbol{m} / \partial T$ afterwards, but keeping in mind its physical origin.

%Similar to the procedure of using the divergent peak in the magnetic susceptibility to locate the transition temperature of the continuous $XY$ model\cite{JFYu}, 
Similar to the procedure used in the continuous $XY$ model to locate the transition temperature\cite{JFYu}, we vary the applied field, and obtain the peak positions of $-\partial \boldsymbol{m} / \partial T$, as presented in Fig. \ref{q5_Tc}(a).
To determine the critical points for a given $D$, an extrapolation to a zero field is performed by a power law fitting
% \begin{equation}
  $T_p-T_c \sim h^x.$
 %\end{equation}
As demonstrated in Fig. \ref{q5_Tc}(a), the results with $D=40$ are obtained as $T_{c1}=0.9038$ and $T_{c2}=0.9557$.
Likewise, we replicate the above process with different bond dimensions, and obtain the converged transition temperatures, i.e. $T_{c1}=0.9063$ and $T_{c2}=0.9557$, as shown in Fig. \ref{q5_Tc}(b). Both agree well with the estimations from other researches\cite{Kumano, Christophe, Chenyong, Borisen2, Chatterjee}.

%
%  \begin{figure}[htbp]
%    \begin{center}
%      \includegraphics[width=0.25\textwidth,clip,angle=0]{q5_Tc_D.eps}
           % Here is how to import EPS art [width=20mm,height=10mm][width=9cm,trim=0.5cm 21cm 0 2cm]
%      \caption{\label{q5_Tc_D}(Color online) The transition temperatures of 5-state clock model versus the tensorial bond dimension to obtain the converged $T_c$.}
%    \end{center}
%  \end{figure}
%

%As shown clearly in Fig. \ref{q5_Tc}(a), two peaks shift oppositely with a varying field. The upper one is similar to the $XY$ case\cite{JFYu}, and the KT transition is well demonstrated, i.e., the stronger a field, the higher the transition temperature, because more heat energy is needed to overcome the additional barrier induced by the symmetry breaking field. However, the lower one behaves oppositely, in a slow and flat way, which may indicate a different scenario. Beside the vortices, we also note that the domain wall is another typical topological excitation in the magnetic systems\cite{Ortiz, Einhorn, Fertig} responsible for the melting of magnetic order.

Once obtaining the critical points, we can investigate the central charge $c$ as well as the critical exponent $\delta$ to determine the universality class of the transitions. 
According to the results of CFT\cite{ Nightingale, Afflect}, we calculate the finite-size partition function on a torus, to obtain the central charge at two critical points and the sandwiched critical phase as $c=1.04$, which indicates that both transitions belong to the same $c=1$ CFT class. Meanwhile, the critical exponent $\delta$ is calculated, which signifies the change of the system magnetization with the applied magnetic field at the transition point as $m\sim h^{1/\delta}$. The results are $\delta_1=15.81$ and $\delta_2=15.77$ respectively, by using the bond dimension $D=70$. Both are consistent well with the theoretical value $\delta=15$ for the KT transition in the 2D $XY$ model\cite{KT2}. Combining $c$ and $\delta$ together, it probably implies two KT-type transitions. As also shown clearly in Fig. \ref{q5_Tc}(a), the upper critical point shifts with the applied magnetic field, the stronger a field, the higher the transition temperature, similar to the $XY$ case\cite{JFYu}, because more heat energy is needed to overcome the additional barrier introduced by the magnetic field.
However, as seen in Fig. \ref{q5_Tc}(a), the lower one moves oppositely, which seems to indicate a different scenario. Besides the vortex excitation, another typical topological excitation responsible for the melting of the magnetic order in magnetic systems is the domain wall\cite{Ortiz, Einhorn, Fertig, Chatterjee}, which probably plays an important role in this transition.

%which seems a little controversial to the results in Fig. \ref{q5_Tc}(a) and Fig. \ref{MC_all}.

To clarify the mechanism, we adopt the procedure of Refs.~[\onlinecite{Soumyadeep, Deng}] and the references therein, to investigate the influence of the vortices excitation on the phase transitions by introducing a parameter $\lambda$ to adjust the vortex core energy as
  \begin{equation}\label{eqH5lamda}
    H = - J \sum_{\left< ij \right>} \cos(\theta_{i}-\theta_{j}) + \lambda\sum_{i'}\left |\omega_{i'}\right |,
  \end{equation}
where $\omega_{i'}=(\delta_{ba}-\delta_{cb}-\delta_{dc}-\delta_{ad})/5$, and $\delta_{ba}$ is $s_b-s_a$ wrapped in $[-1, 1]$. $s_a, s_b, s_c, s_d$ are spins on four vertexes of a square plaquette labelled by $i'$.

  \begin{figure}[htbp]
    \begin{center}
      \includegraphics[width=0.48\textwidth,clip,angle=0]{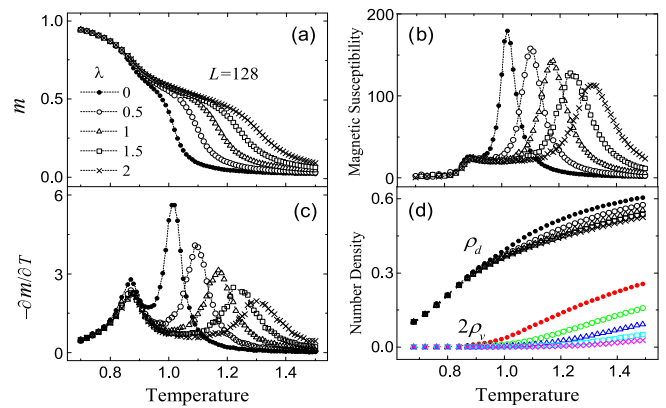}
           % Here is how to import EPS art [width=20mm,height=10mm][width=9cm,trim=0.5cm 21cm 0 2cm]
      \caption{\label{MC_all}(Color online) MC simulation of the Hamiltonian (Eq. \ref{eqH5lamda}) with $L=128$ for different $\lambda$: (a) magnetization; (b) magnetic susceptibility; (c) $-\partial \boldsymbol{m} / \partial T$; (d) number densities of the domain walls ($\rho_d$) and the vortices ($\rho_v$), where $\rho_v$ is multiplied by 2 for a better view.}
    \end{center}
  \end{figure}

By MC simulations about the above Hamiltonian (Eq. \ref{eqH5lamda}) on a square lattice with $L=128$, we obtain the magnetization, the magnetic susceptibility and the deduced $-\partial \boldsymbol{m} / \partial T$ for different $\lambda$, as all shown in Fig. \ref{MC_all}. Increasing the vortex core energy to suppress its formation, a clear shift of the upper critical point can be seen from each curve.
Again, $-\partial \boldsymbol{m} / \partial T$ looks much more convincing than the magnetic susceptibility for the lower transition.
More importantly, as manifested in Fig. \ref{MC_all}(b) and (c), this lower temperature phase transition is barely affected by the vortex suppression, which strongly suggests it is dominated by the domain wall excitation\cite{Ortiz, Einhorn, Fertig, Chatterjee}.
A more intuitive illustration is presented in Fig. \ref{MC_all}(d), i.e. the number density of each excitation, by adopting the definition in Ref.~[\onlinecite{Soumyadeep}].
One can clearly observe that, near the lower transition point, the number density of the domain wall decreases negligibly, while the vortices are greatly suppressed even eliminated, when increasing $\lambda$.

Furthermore, we calculate the aforementioned universal entropy $\ln{g}$ of CFT on a Klein bottle\cite{TuPRL} at the critical points, because CFT asserts the two transitions in this model are KT-type but with different $g$\cite{TuPrivate}. Our computation gives $g_1=3.30$ and $g_2=3.09$ respectively, both of which agree well with the CFT conclusion\cite{TuPrivate}. From the foregoing discussions, we can conclude that both transitions are indeed KT-type, but with subtle differences: the upper one is attributed to the unbinding of the vortices pairs, while the lower one is dominated by the domain wall excitation instead; and they belong to different CFTs.
The differences may be closely related to why the magnetic susceptibility works fine for the upper transition but not so well for the lower one; and why the transition points shift oppositely with the external field as shown in Fig. \ref{q5_Tc}(a). They may also be the reason why studies from different groups would give controversial estimations about the nature of the transitions.

%From Fig. \ref{q5_Tc}(a) and Fig. \ref{MC_all}, we can conclude that the upper transition is indeed KT-type, conforming to the vortex anti-vortex pair unbinding scenario. While, the lower one seems different, wherein the domain walls play a dominant role instead of the vortices. According to a CFT in Refs.~[\onlinecite{ Nightingale, Afflect}], we calculate the finite-size partition function on a torus, to obtain the central charge at both critical points and the critical phase in-between as $c=1.04$, which indicates that both transitions belong to the same $c=1$ CFT class. Meanwhile, the critical exponent $\delta$ is computed, which signifies the change of the system magnetization with an applied field at the critical points. They are $15.81$ and $15.77$ respectively by using the bond dimension $D=70$. Both are well consistent with the theoretical value $\delta=15$ for the KT transition of 2D $XY$ model\cite{KT2}. Considering $c$ and $\delta$ together, it obviously denotes two KT-type transitions, which seems a little controversial to the results in Fig. \ref{q5_Tc}(a) and Fig. \ref{MC_all}. So further, we compute the aforementioned universal entropy $g$ of a CFT on a Klein bottle\cite{TuPRL} at the critical point, because for this model the CFT proposes two KT-type transitions with different $g$\cite{TuPrivate}. Our computation gives $g_1=3.30$ and $g_2=3.09$ respectively, both of which are close to the CFT expectations\cite{TuPrivate}, confirming that both transitions are indeed KT-type, but with subtle differences.

  \begin{figure}[htbp]
    \begin{center}
      \includegraphics[width=0.48\textwidth, clip, angle=0]{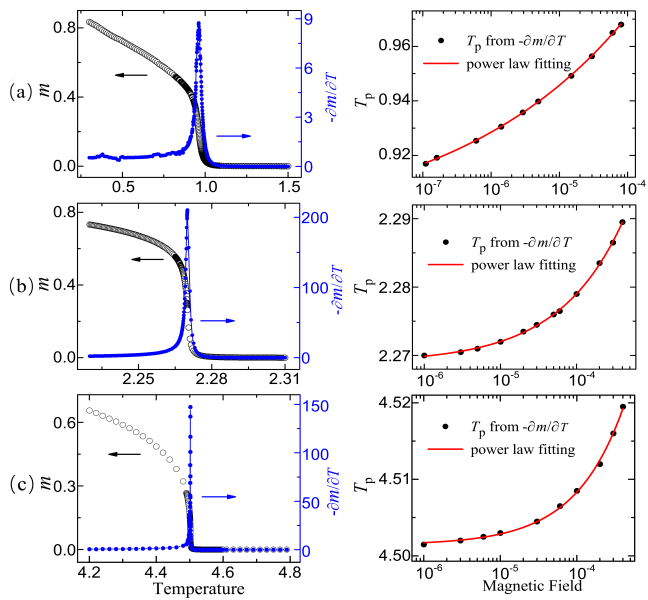}
      \caption{\label{combine}(Color online) $-\partial\boldsymbol{m} / \partial T$ and the power law fitting of its peak position varying with the applied field: (a) 2D $XY$ model with $D=40$; (b) 2D Ising model with $D=40$; (c) 3D Ising model with $D=10$.}
    \end{center}
  \end{figure}

%  \begin{figure}[htbp]
%    \begin{center}
%      \includegraphics[width=0.45\textwidth,clip,angle=0]{Ising_2D.eps}
           % Here is how to import EPS art [width=20mm,height=10mm][width=9cm,trim=0.5cm 21cm 0 2cm]
%      \caption{\label{Ising_2D}(Color online) HOTRG results of the classic Ising model on the square lattice with $D=40$: (a) magnetization and susceptibility; (b) peak of the magnetic susceptibility, and the power law fitting; (c) magnetization and $-\partial \boldsymbol{m} / \partial T$; (d) peak of $-\partial \boldsymbol{m} / \partial T$ and the power law fitting.}
%    \end{center}
% \end{figure}

Briefly, with an auxiliary external magnetic field, the function $\partial \boldsymbol{m} / \partial T$ accurately reflects the interplay of the field and the temperature, and captures the implicit dynamics of the excitations in this model, hence correctly describes the phase transitions.
% and their precise locations, though supplementary measurements may be requisite to precisely determine the universality class of a transition.
%That is because it originates from the fundamental free energy $F$, which includes all possible configurations of the system in thermodynamical equilibrium, such as the multiple topological excitations and the potential competition/interaction among them.
While, the derivative of the free energy $F$ with respect to each single parameter, such as the specific heat or the magnetic susceptibility, is inadequate for lacking of information about the internal competition/interplay among those mingled complex excitations.
%to detect the exotic phase transition with mingled complex excitations, because of lacking information about the interplay among them.
The auxiliary magnetic field and the cross derivative provide us a convenient way to observe the response/dynamics of different excitations.

%Some may argue the physical meaning or the validity of making use of this second-order partial derivative of the free energy. While, we can imagine slicing the 3-dimensional curve $F(T, h)$ along $h$-axis, and performing the derivative $\partial{F(T, h)}/{\partial{T}}$ for each field slice $h$, then observe its evolution along different $h$, namely the second derivative mentioned above; or equivalently slicing $F(T, h)$ along $T$-axis and getting $\partial{F(T, h)}/{\partial{h}}$, then investigate its evolution along different $T$, either can thus delicately describe the interplay between the external magnetic field and the temperature, and deduce the dynamic properties of the system, such as the information about the phase transitions.

%  \begin{figure}[htbp]
%    \begin{center}
%      \includegraphics[width=0.45\textwidth,clip,angle=0]{Ising_3D.eps}
           % Here is how to import EPS art [width=20mm,height=10mm][width=9cm,trim=0.5cm 21cm 0 2cm]
%      \caption{\label{Ising_3D}(Color online) HOTRG results of the classic Ising model on the cubic lattice with $D=10$: (a) magnetization and susceptibility; (b) peak of the magnetic susceptibility, and the power law fitting; (c) magnetization and $-\partial \boldsymbol{m} / \partial T$; (d) peak of $-\partial \boldsymbol{m} / \partial T$ and the power law fitting.}
%    \end{center}
%  \end{figure}

To further check the universality of this idea, we apply it to the 2D $XY$, the 2D Ising, and the 3D Ising models separately. A sample of $-\partial\boldsymbol{m} / \partial T$ and the power law fitting of the peak position for each model are illustrated in Fig. \ref{combine} as (a), (b), and (c) respectively. 
For the $XY$ case, the transition temperature is obtained at $T_c=0.8924(16)$, which is coincident to the previous estimation\cite{JFYu} from the magnetic susceptibility $T_c=0.8921(19)$ with same bond dimension $D=40$, and both conform to the results from other methods like MC\cite{Hasenbusch2, Tomita} at $T_c=0.89294(8)$.
For the 2D Ising case, the power law extrapolation yields the transition temperature at $T_c=2.26893(18)$, and a simultaneous prediction from the magnetic susceptibility (not shown in the figure) is $T_c=2.26904(22)$. They agree well with each other, and with the exact value $T_c=2/\ln{(\sqrt{2}+1)}\sim2.26919$, even using a relatively small bond dimension $D=40$.
As to the 3D Ising case, the same procedure is carried out with the bond dimension $D=10$. The similar efficiency of the function $-\partial \boldsymbol{m} / \partial T$ is clearly demonstrated once again, from which the critical temperature is located at $T_c=4.5014(2)$. Also, the $T_c$ is determined at $4.5013(1)$ from the magnetic susceptibility. Both are consistent with the prediction at $T_c=4.5015$ by the HOTRG calculation with same $D$\cite{Xie2}.
What's more, we can also observe the singularity in the $-\partial \boldsymbol{m} / \partial T$ curve of the 2D/3D Ising model, indicating a second-order phase transition. 
It becomes more manifest and sharper if lowering the field down to zero, then a direct and accurate determination of the critical point can be obtained with no need for an extrapolation.

%As to the $XY$ model and the 5-state clock model discussed in Fig. \ref{fe}(b), much shorter peaks imply less singularity, though a quite sharp decline still remains across the transition point. The transition seems higher than second order.

%  \begin{figure}
%    \begin{center}
%      \includegraphics[width=0.45\textwidth,clip,angle=0]{xy.eps} \caption{\label{xy}(Color online) HOTRG results of the $XY$ model on the square lattice with $D=40$: (a) magnetization and $-\partial \boldsymbol{m} / \partial T$; (b) peak of $-\partial \boldsymbol{m} / \partial T$ and the power law fitting.}
%    \end{center}
%  \end{figure}

These examples have verified the capability of our idea perfectly, and the proposed cross derivative $\partial^2{F}/{\partial{T}\partial{h}}$ seems more versatile and effective, no matter a transition is trivial or exotic, especially when multiple exotic excitations are involved and other quantities/methods are difficult to clarify. Also, we think this strategy is universal, as long as the free energy can be calculated accurately with a weak external magnetic field included.
Experimentally, one can measure the system magnetization $\boldsymbol{m}(T, h)$, from which the phase transition information can be easily deduced.
More importantly, the magnetic field and the magnetization in Gibbs free energy or the Hamiltonian are just one typical conjugate pair of generalized force and displacement\cite{Reichl}. Likewise, other conjugate pairs, if introduced into the Hamiltonian to regulate a system's behavior, would play a similar role in investigating the phase transitions, e.g. the electric field and the polarization in an electronic system, which could be integrated into formula \eqref{formu2} similarly.
Thus, this idea will greatly enrich our vision and means to study the phase transitions both theoretically and experimentally.

%\section{Summary}\label{Summary}
Considering its accuracy and simplicity, the idea we proposed in this work is efficient and universal to investigate the phase transitions in classical spin systems, trivial or complex, 2D or 3D. The predictions will be more accurate if the free energy or the physical quantities involved could be computed more precisely.

%Acknowledgement ...
We are grateful to Hong-Hao Tu, Fuxiang Li, and Yu-Chin Tzeng for valuable discussions and comments. Y. Chen thanks Mr. Yuan Si for helps on Monte Carlo simulations. This work was supported by the Shanghai Pujiang Program (No. 17PJ1407400), the National Natural Science Foundation of China (No. 11774420), the National R\&D Program of China (No. 2016YFA0300503, No. 2017YFA0302900), and the Natural Science Foundation of Hunan Province (No. 851204035).

%\vspace{0.25cm}

%\noindent{Corresponding authors:}\\
%Corresponding authors:
%\noindent{{yujifeng@hnu.edu.cn}}\\
%\noindent{{qingtaoxie@ruc.edu.cn}}

%\bibliographystyle{apsrev4-1}
\bibliography{classical_spin_models}

\end{document}